\documentclass[12pt]{article}
\ifx\pdfoutput\undefined
\usepackage[dvips,bookmarks]{hyperref}	
\else
\usepackage{hyperref}	
\fi
\hypersetup{colorlinks,bookmarksopen,bookmarksnumbered,citecolor=blue,
	pdfstartview=FitH}
\def\hhref#1{\href{http://arxiv.org/abs/hep-th/#1}{hep-th/#1}} 
\def\mhref#1{\href{mailto:#1}{#1}}		
\catcode`\|=\active 
\def|{\relax\ifmmode\delimiter"026A30C \else$\mathchar"026A$\fi}
\def\slap#1#2{\setbox0=\hbox{$#1{#2}$}
	#2\kern-\wd0{\hbox to\wd0{\hfil$#1{/}$\hfil}}}
\def\sla#1{\mathpalette\slap{#1}}
\def\on#1#2{{\buildrel{\mkern2.5mu#1\mkern-2.5mu}\over{#2}}}
\catcode`@=11
\def\secteqno{\@addtoreset{equation}{section}%
\def\theequation{\thesection.\arabic{equation}}}
\catcode`@=12
\secteqno
\newcommand{\be}{\begin{equation}}
\newcommand{\ee}{\end{equation}}
\newcommand{\bea}{\begin{eqnarray}}
\newcommand{\eea}{\end{eqnarray}}
\def\bref#1{(\ref{#1})}
\newcommand{\nn}{\nonumber}
\begin{document}
\thispagestyle{empty}
\vfill
\hfill November 20, 2002\par
\hfill KEK-TH-855, YITP-SB-02-72\par
\vskip 20mm
\begin{center}
{\Large\bf A new holographic limit of AdS$_5\otimes$S$^5$}\\
\vskip 6mm
\medskip

\vskip 10mm
{\large Machiko\ Hatsuda$^\ast$~and~Warren\ Siegel$^\star$}\par
\parskip .15in
{\it $^\ast$Theory Division,\ High Energy Accelerator Research Organization (KEK),\\
\ Tsukuba,\ Ibaraki,\ 305-0801, Japan} \\
{\small e-mail:\ \mhref{mhatsuda@post.kek.jp}} \\
\parskip .15in
{\it 
$^\star$C.N. Yang Institute for Theoretical Physics,\\ 
 State University of New York,\ Stony Brook,\ NY \ 11794-3840, USA}\\
{\small e-mail:\ \mhref{siegel@insti.physics.sunysb.edu} } 
\end{center}

\baselineskip .23in
\begin{abstract}
We re-examine the projective lightcone limit of the gauge-invariant 
Green-Schwarz action on 5D anti-de Sitter $\otimes$ the five-sphere.
It implies the usual holography for AdS$_5$, but also
(a complex) one for S$^5$.
The result is N=4 projective superspace,
which unlike N=4 harmonic superspace can describe 
N=4 super Yang-Mills off shell.
\end{abstract} 
\par\par
\newpage
\setcounter{page}{1}
\parskip=7pt
\section{ Introduction}\par

The anti-de Sitter/conformal field theory correspondence
is an interesting attempt to relate string theory to
quantum chromodynamics \cite{[1]}.  
The states of the (Type IIB) string are identified as 
color-singlet bound states of Yang-Mills
(N=4 supersymmetric, in the simplest version).
Although the equivalence is still conjectural,
its study has led to new insights on both theories.
It is based on the assumption of holography, that one dimension of AdS$_5$
becomes irrelevant, so that the usual four dimensions of spacetime can
be obtained.  Holography has not been obtained dynamically, but is
imposed as part of the definition of calculated quantities, and can be
interpreted as an expansion about the boundary of AdS$_5$, whose
convergence is yet to be determined.

The conventional correspondence identifies the Yang-Mills fields
with those of a dual (Type I) open string whose ends are confined
to the boundary (branes).  Hence these fields are not seen explicitly
in this approach.  However, in string theory closed strings are not true
bound states of open strings, but appear kinematically in the free theory,
in a manner similar to bosonization/fermionization in free two-dimensional
theories \cite{[2]}.  (They appear in one-loop graphs, not through ladder graphs
representing multiple exchanges.)  In a previous paper an alternative
correspondence was proposed \cite{[3]}, which identifies the Yang-Mills fields
with the matrix fields of the random-lattice approach to string theory,
where planar Feynman diagrams of partons are identified with lattice
discretizations of the string worldsheet path-integral through the
1/N expansion \cite{[4]}.  

That paper also proposed a different approach to holography, based on
treating anti-de Sitter space through an expansion about the projective
lightcone (its zero-radius limit), the original higher-dimensional
formulation of conformal symmetry \cite{[5]}.  In that limit the fifth dimension
becomes nondynamical; effectively, rather than restricting the theory to the
boundary, the boundary has been expanded into the whole space.
When combined with the random lattice approach, the result is that
in the path integral the fifth coordinate acts as Schwinger parameters
(a fifth, ``proper-time" coordinate), converting the Gaussian propagators
of the partons in the usual random lattice approach into those of
conventional field theory.

In that paper calculations were simplified by taking
the projective lightcone limit only after choice
of a special gauge that is not applicable to the general case \cite{[6]}.
Another consequence was that the gauge-invariant form of the
limiting action was not examined.  Here we take the limit before gauge
fixing.  While AdS$_5$ is contracted to the projective lightcone,
S$^5$ is not contracted to a point, but also to a flat four-dimensional space,
as expected from symmetry with AdS$_5$ under Wick rotation.
Including the related contraction of the fermionic coordinates,
the resulting superspace is that appropriate to the N=4 projective
superspace description of off-shell N=4 super Yang-Mills, a direct
generalization of N=2 projective superspace \cite{[7]}.  
The modified reality conditions of superfields in this approach
follow directly from this Wick rotation.
(Such superspaces
have also been applied to the harmonic superspace approach \cite{[8]},
but the result was always on shell \cite{[9]}, because the boundary conditions
for the coordinates of the internal R symmetry differ in the two
approaches \cite{[10]}.)

The resulting random lattice action for the superparton is
identical to the usual Casalbuoni-Brink-Schwarz superparticle action
(for D=4, N=4) after a simple redefinition, and thus describes 
4D N=4 super Yang-Mills (as known, e.g., from lightcone quantization).
(In \cite{[3]} this action was obtained after gauge fixing.)
However, the form of the action before the redefinition directly
implies projective superspace because of the appearance of the
internal coordinates with derivatives, and half of the fermionic
variables as Lagrange multipliers (i.e., without derivatives).
The choice of off-shell superspace, and in particular the appearance
of internal coordinates, thus depends on the  
treatment of the second-class constraints. 
(The on-shell superspace is always the same, up to trivial
Fourier transformation in fermionic coordinates,
because the physical spectrum is the same.)
The appearance of internal coordinates also allows more
general gauges, and the use of the infinite number of
auxiliary fields required for off-shell N=4 supersymmetry.

The equations we give below as ``constraints", when applied in the 
harmonic formalism, imply the equations of motion (Klein-Gordon, Weyl, 
or Maxwell equations) due to the condition of regularity in the 
harmonic coordinates, as imposed by the harmonic expansion.  On the 
other hand, in the projective formalism these constraints simply 
restrict dependence of the fields to the coordinates of projective 
superspace, which have a Laurent expansion in the projective 
coordinates.

This has already been demonstrated for the case of the N=2 scalar 
multiplet:  In the harmonic case, only a subset of the constraints 
(``analyticity") was imposed as constraints, the remainder were used as 
field equations.  In the projective case, all were imposed, and solved 
off-shell, and the field equations following from the action were those 
equations called ``field equations" below.

In the case of N=3 super Yang-Mills, only the harmonic case has been 
treated \cite{N3}, and the same separation was made.  However, quantization of 
this formalism proved difficult, and a Fermi-Feynman gauge was not 
found \cite{DdMc}.  Since the whole point of an off-shell formalism is 
quantization (otherwise one can simply look at solutions of the 
classical field equations), and the point of gauge invariance is to be 
able to choose arbitrary gauges (since different choices have different 
advantages, and gauge independence is part of unitarity), this theory 
still needs a formulation where Feynman rules for general gauges can be 
written.  Furthermore, since this theory has an N=4 supersymmetry, 
ideally an N=4 superspace solution is desired.  However, so far it has 
not been possible to give an off-shell N=4 harmonic superspace 
formulation of this theory.

The AdS/CFT approach has shown promise in giving a superspace approach 
to this theory.  It naturally leads to an N=4 formulation, and has been 
used previously in conjunction with on-shell N=4 harmonic superspace.  
However, as we will see below, manifesting as much superconformal 
invariance as possible leads naturally to projective superspace, as a 
direct supersymmetrization of the coset space approach used both for 
the standard representation of the conformal group on spacetime and for 
the AdS Green-Schwarz superstring action.  In fact, the above limiting 
procedure leads directly to a classical mechanics action for N=4 super 
Yang-Mills on projective superspace.  The projective approach solves 
the problem that prevented an N=4 formulation (in both harmonic and 
earlier approaches), by solving the constraints without going on shell. 
  Although at this point we have quantized (and derived a field theory 
action) from this mechanics action only for the lightcone gauge (where 
it trivially reduces to the usual lightcone methods), these results are 
encouraging in their avoidance of earlier problems and relation to 
previous useful results, as well as their direct relation to the 
AdS/CFT correspondence.

\section{ Projective superspaces for D=4}

\subsection{ Coset spaces} 

There is a general kind of coset-space construction that appears frequently
in physics, but the relation of the various cases is never described.
We will call them ``half-coset" spaces because that describes both
our construction and notation for them.

The general construction is:  (1) Given a group G, consider its Cartan
subalgebra, and pick from it one generator, perhaps a linear combination
of the natural basis chosen.\break
(2) Divide up the generators of G, in a basis of eigenvectors of the chosen
U(1) generator, into those with positive eigenvalues, G$_+$
(raising operators), those with negative, G$_-$ (lowering operators), 
and those with vanishing eigenvalue, G$_0$ (which
includes the Cartan subalgebra, and in particular the chosen U(1) generator,
but in general will be nonabelian).
(3) The subalgebra of interest is that generated by G$_+$.  For notational
and pedagogical purposes, it will usually be convenient to label this
subalgebra by the half-coset notation as G/G$_0$+, since G and G$_0$
will be more readily identifiable.  It can also be interpreted as the coset
with ``isotropy" subgroup the semidirect product of G$_0$ and G$_-$.

We first give some examples from string theory.
(In the following subsections we describe cases more closely
related to the present paper.)
One familiar example is Gupta-Bleuler quantization of
string theory.   There G is the (super)Virasoro algebra, 
G$_0$ its zero-modes, and G$_+$ creates
unphysical states from the vacuum, 
or from physical states gauge-equivalent ones.
A similar method can be applied to general groups of constraints
in Gupta-Bleuler quantization \cite{[11]}.
	
A case of recent interest is the ``pure spinor", which appears in Berkovits'
description of the superstring \cite{[12]}.  It has been described as the coset
SO(10)/U(5), but more accurately it is SO(10)/U(5)+, where the relevant U(1)
is unambiguous in U(5)=SU(5)U(1).  (We omit the direct product symbol
``$\otimes$" in group products, just as the usual product symbol
``$\times$" is conventionally omitted in algebra.)
	
Another example, closely related algebraically to the ones relevant
to this paper, 
where such a construction is implicit, is the manifestly
T-dual formulation of the massless sector of oriented closed strings
(graviton, axion, dilaton) \cite{[13]}.  There the half-coset is
GL(2D)/GL(D)$^2$+ in D dimensions, where the U(1) (actually GL(1)),
from\break
GL(D)=SL(D)GL(1), is not the diagonal one (which cancels the one
from the GL(2D)).  If we divide the defining-representation group
element, a (2D)$^2$ matrix, into four equal quadrants, the two diagonal
quadrants give G$_0$, while the lower-left quadrant gives G$_+$,
which is Abelian in this case.  The G group element represents the
fields of the theory (less the dilaton); in the gauge where G$_0$ and
G$_-$ have been used to gauge away three quadrants, the symmetric part
of G$_+$ corresponds to the metric and the antisymmetric part to the
axion two-form.  Previously this had been described in terms of two
of the adjacent quadrants; that non-coset description follows from this one
if G$_-$ and half of G$_0$ are used to gauge away the other half of the
matrix.  The resulting D$\times$2D matrix still represents GL(2D) by
multiplication on the longer side, and one of the local GL(D)'s on the other.
In that approach the ratio of the two surviving quadrant matrices 
(a ``projective" space) is gauge
invariant under the remaining gauge symmetry, which is equivalent to
using that remaining symmetry to gauge away all but G$_+$.
(Here we refer to only the non-derivative, ``tangent-space" gauge symmetry;
for this particular case there remains the usual derivative gauge
transformations of the metric and axion field.)
Note that in this case, where G is a GL group, the advantage of working 
directly with group elements instead of algebra elements is that there
is no need to exponentiate \cite{[14]}.

\subsection{ Superconformal cosets} 

Probably the most familiar use of such a construction is for the 
case of the usual
coordinate representation of the conformal group (for spin zero).
There the half-coset is \break
SO(D,2)/SO(D$-$1,1)SO(1,1)+.  The U(1) is the
scale generator, the rest of G$_0$ is the Lorentz generators, G$_-$ is
conformal boosts, and G$_+$ is translations, corresponding to the
usual spacetime coordinates.

In D=4, if we replace these groups with their covering groups, the
half-coset is SU(2,2)/SL(2,C)GL(1)+.  For our purposes it will be more
convenient to consider the Wick rotation, the covering of
SO(3,3)/SO(2,2)SO(1,1)+, which is GL(4)/GL(2)$^2$+, where we have
thrown in an extra GL(1) on top and bottom for convenience.

The procedure is then the same as for T-dual axionic gravity as
described above (but now we deal with coordinates rather than fields).
The resulting representation of the conformal group is the same as
with orthogonal groups, but in spinor notation:  It is the ``quaterionic
projective" version of ADHM twistors \cite{[15]} (again if we identify by using 
half the matrix).

Such representations of conformal groups have been generalized to
superconformal groups \cite{[16]}, and in particular for chiral \cite{[17]} and other \cite{[9]}
``subsuperspaces".  For our purposes, it will be sufficient to consider
the cases GL(4|N)/GL(2|n)GL(2|N-n)+, which are direct generalizations
of the bosonic (N=n=0) case.  Now only the two diagonal blocks need be 
square; e.g., n=0 describes chiral superspace.
However, the cases N=2n are actually the most relevant ones,
since they describe CPT self-conjugate multiplets (the N=2 scalar
multiplet, N=4 Yang-Mills, and N=8 supergravity).
In general these half-cosets are Abelian subgroups:
These subspaces of the full superspace have no torsion.

\subsection{ Superspaces}

Now we consider these spaces GL(4|N)/GL(2|n)GL(2|N-n)+
in more detail, and the description of supersymmetric theories
on them.  We divide up the indices as
\bea {\cal A} = A, A' \eea
where $\cal A$ is the (4|N) index, $A$ is (2|n), and $A'$ is (2|N-n).
We then separate out the bosonic (2-component Weyl spinor) 
and fermionic (internal) indices as
\bea
 A = \alpha, a; \quad A' = \dot\alpha, a' 
 \eea
Writing the G superspace coordinates (GL(4|N) group element
in the defining representation) as
$z_{\cal A}{}^{\cal M}$ and its inverse as $z_{\cal M}{}^{\cal A}$, 
where the superconformal group generators $G_{\cal M}{}^{\cal N}$ 
act on the index $\cal M$, and thus the covariant derivatives 
$D_{\cal A}{}^{\cal B}$ act on the other side, on the index $\cal A$, 
we have
\bea
 G_{\cal M}{}^{\cal N} = z_{\cal A}{}^{\cal N}\partial_{\cal M}{}^{\cal A},
\quad D_{\cal A}{}^{\cal B} = z_{\cal A}{}^{\cal M}\partial_{\cal M}{}^{\cal B} 
\eea
with the usual implicit sign factors for ordering of fermionic indices,
where $\partial_{\cal M}{}^{\cal A}=\partial/\partial z_{\cal A}{}^{\cal M}$.
The two are then linearly related as
\bea
 D_{\cal A}{}^{\cal B} = z_{\cal A}{}^{\cal M}z_{\cal N}{}^{\cal B}
G_{\cal M}{}^{\cal N} ~~.
\eea

We now consider the steps necessary to
restrict the coordinates $z_{\cal A}{}^{\cal M}$ 
that appear as arguments of the field to the half-coset.  We begin
by imposing the constraints (which also generate the gauge
transformations of the isotropy group)
\bea
 D_A{}^B = D_{A'}{}^{B'} = D_{A'}{}^B = 0 \eea
on the field.  
(This will define the superspace.
Below we will consider the more general case of
superfields with indices, where $D_A{}^B$ and $D_{A'}{}^{B'}$
will be set to equal matrix representations,
which will lead to the same coordinates,
but add spin operators to the group generators.)
This leaves covariant derivatives 
\bea
 D_A{}^{B'} = p_\alpha{}^{\dot\beta}, d_\alpha{}^{b'},
\bar d_a{}^{\dot\beta}, t_a{}^{b'} \eea
corresponding to the surviving coordinates.

We now solve the constraints explicitly, starting with
\bea
 0 = D_{\cal A}{}^B = z_{\cal A}{}^{\cal M}\partial_{\cal M}{}^B
\quad\Rightarrow\quad \partial_{\cal M}{}^A = 0 
\eea
explicitly eliminating dependence on $z_A{}^{\cal M}$.
(At this point we have the ``half-matrix" mentioned above
that appeared in earlier approaches to some theories.)
The remaining constraint $D_{A'}{}^{B'}=0$ then says that the field
is invariant under general linear transformations on the $A'$ index,
so that the remaining coordinates $z_{A'}{}^{\cal M}$ can appear
in the field only in the combination
\bea
 w_{M'}{}^N \equiv (z_{A'}{}^{M'})^{-1}z_{A'}{}^N 
 \eea
where ``$(z_{A'}{}^{M'})^{-1}$" means to take only the inverse of
the matrix $z_{A'}{}^{M'}$ and not the corresponding part of the
inverse of $z_{\cal A}{}^{\cal M}$.  
These can be separated into
\bea
 w_{M'}{}^N = x_{\dot\mu}{}^\nu, \theta_{m'}{}^\nu, 
\bar\theta_{\dot\mu}{}^n, y_{m'}{}^n 
\eea
where $x$ are the usual spacetime coordinates,
$\theta$ and $\bar\theta$ are half of the anticommuting coordinates
of the full superspace, and $y$ are the internal coordinates.
When acting on fields
depending on only $w$, the superconformal generators take the 
form, using the notation 
$\partial_M{}^{N'}\equiv\partial/\partial w_{N'}{}^M$,
\bea
 G_M{}^{N'} = \partial_M{}^{N'}, \quad
G_M{}^N = w_{P'}{}^N\partial_M{}^{P'}, \quad
G_{M'}{}^{N'} = -w_{M'}{}^P\partial_P{}^{N'}, 
\eea
\bea
 G_{M'}{}^N = -w_{M'}{}^P w_{Q'}{}^N\partial_P{}^{Q'} ~~~.\eea

However, while the symmetry generators $G_{\cal M}{}^{\cal N}$
depend only on the surviving coordinates and their derivatives
(i.e., those coordinates are a realization of the full group),
the remaining covariant derivatives, while containing derivatives
with respect to only the (half-)coset coordinates, can have coordinate 
dependence on some of the coordinates of the isotropy group.
For example, in chiral superspace, in the chiral representation,
the supersymmetry generators contain no $\bar\theta$'s nor
$\partial/\partial\bar\theta$'s, but the surviving spinor covariant
derivative does have a $\bar\theta$ term.
To simplify the resulting expressions it is useful to extend the
change of variables from $z$ to $w$ as represented by the
matrix decomposition
\bea
 \pmatrix{ z_A{}^M & z_A{}^{M'} \cr z_{A'}{}^M & z_{A'}{}^{M'} \cr}
= \pmatrix{ I & v \cr 0 & I \cr} \pmatrix{ u & 0 \cr 0 & u' \cr} 
\pmatrix{ I & 0 \cr w & I \cr} 
\eea
or
\bea
 z_{A'}{}^{M'} = u_{A'}{}^{M'}, \quad
z_{A'}{}^M = u_{A'}{}^{N'}w_{N'}{}^M, \quad
z_A{}^{M'} = v_A{}^{B'} u_{B'}{}^{M'}, \eea
\bea
 z_A{}^M = u_A{}^M + v_A{}^{B'} u_{B'}{}^{N'}w_{N'}{}^M ~~.\eea
Later we will need also the inverse
\bea
 \pmatrix{ z_M{}^A & z_M{}^{A'} \cr z_{M'}{}^A & z_{M'}{}^{A'} \cr}
= \pmatrix{ I & 0 \cr -w & I \cr} \pmatrix{ u^{-1} & 0 \cr 0 & u'^{-1} \cr} 
\pmatrix{ I & -v \cr 0 & I \cr} \eea
\bea z_M{}^A = u_M{}^A, \quad
z_{M'}{}^A = -w_{M'}{}^N u_N{}^A, \quad
z_M{}^{A'} = -u_M{}^B v_B{}^{A'}, \eea
\bea
 z_{M'}{}^{A'} = u_{M'}{}^{A'} + w_{M'}{}^N u_N{}^B v_B{}^{A'} \eea
where $u_M{}^A$ is the matrix inverse to $u_A{}^M$, etc.
($u$ and $u'$ now act as the vielbeins for GL(2|n) and GL(2|N-n).)
In terms of these the surviving covariant derivatives
act on the fields as
\bea
 D_A{}^{B'} = u_A{}^M u_{N'}{}^{B'} \partial_M{}^{N'} \eea

\subsection{ Free field equations}

The free superconformal equations of motion 
(field equations for field strengths) 
on general superspaces can be written
as either \cite{[18]}
\bea
G_{[\cal M}{}^{[\cal N}G_{\cal P)}{}^{\cal Q)} - traces = 0 \quad or\quad
D_{[\cal A}{}^{[\cal B}D_{\cal C)}{}^{\cal D)} - traces = 0 \eea
using graded antisymmetrization $[\phantom{n})$, where 
$z_{\cal A}{}^{\cal M}$ can be used to convert between the two forms.
(A familiar example is that a field equation common to all massless
supersymmetric theories, the generator of $\kappa$ symmetry, can
be written as either $\sla pd=0$ or $\sla pq=0$ in linear combination
with $p^2=0$.)  These equations are those related to the masslessness
condition $p^2=0$ by superconformal transformations.

On projective superspaces the field equations remaining after
imposing the (isotropy) constraints are
\bea
 D_{[A}{}^{[B'}D_{C)}{}^{D')} = 0 ~~.\label{220}\eea
These can be expanded as 
(using antisymmetrization in spinor indices to contract them),
in order of decreasing (scale) dimension,
\bea p^2 = 0 \label{pp} \eea
\bea \sla p d = \sla p \bar d = 0 \eea
\bea (d^2)^{(a'b')} = (\bar d^2)_{(ab)} 
= d_\alpha{}^{b'}\bar d_a{}^{\dot\beta} - p_\alpha{}^{\dot\beta}t_a{}^{b'} = 0 \eea
\bea t_a{}^{(b'}d_\alpha{}^{c')} = t_{(a}{}^{b'}\bar d_{c)}{}^{\dot\beta} = 0 \eea
\bea t_{(a}{}^{(b'}t_{c)}{}^{d')} = 0 ~~.\label{tt}\eea
Field equations of higher dimension follow from those of 
lower dimension by commutation with the constraints;
this corresponds to expansion in the isotropy coordinates.
In fact, such an expansion of the lowest-dimension field equation 
yields the corresponding equations in terms of the generators of
the half-coset subgroup:
\bea G_{[M}{}^{[N'}G_{P)}{}^{Q')} = 0 \quad\Rightarrow\quad
\partial_{[M}{}^{[N'}\partial_{P)}{}^{Q')} = 0 ~~.\label{226}\eea

For the harmonic superspace approach, these constraints alone are enough
to put the cases N=2n=2 \cite{[8]} or 4 \cite{[9]} on shell.  
(That approach for N=2 uses one of these constraints as a field equation instead.
There is also an off-shell harmonic N=3 formalism of  
N=4 \cite{N3}, but it lacks the explicit N=4 superspace used in treatments of  
AdS/CFT.  In reference \cite{CKR} the AdS/CFT representation of the superconformal  
group was gauge fixed and reduced to N=2, but since no dynamics were  
considered, as either string or field theory action or field equations,  
and the result was gauge dependent, it is impossible for them to  
distinguish between N=2 harmonic and projective superspaces.)
But in the projective
superspace approach for N=2n=2, these constraints only define the space
(off shell), and the field equations for the scalar (``arctic") multiplet,
whose field strength (the field itself) is projective, are those given above.
These equations
also describe the free part of N=4 super Yang-Mills for N=2n=4,
although a field theory action has not yet been written.
This can be seen easily by Taylor expanding the superfield in $y$
(the definition of the projective boundary conditions), and in $\theta$.
A simpler (covariant) way is to use supertwistors \cite{[19]}:
Since the equations in terms of the generators involve only
partial derivatives, Fourier transformation gives the algebraic solution
\bea
 G_M{}^{N'} = \zeta_M \zeta^{N'} \quad\Rightarrow\quad \Phi (w) = 
\int d\zeta\ exp \left(w_{M'}{}^N \zeta_N \zeta^{M'}\right) \chi (\zeta) +h.c. \label{stw}
\eea
where the components of $\zeta$ with spinor indices are the usual
Penrose twistors and the rest are fermions, corresponding to the 
$\theta$'s of lightcone superfields (giving the usual helicity expansion).

The on-shell form of the superconformal generators
(acting on $\chi$) is then
\bea
 G_M{}^{N'} = \zeta_M \zeta^{N'}, \quad
G_M{}^N = \zeta_M \partial^N, \quad
G_{M'}{}^{N'} = \zeta^{N'}\partial_{M'}, \quad
G_{M'}{}^N = \partial_{M'} \partial^N 
\eea
where 
$$ [ \partial_{M'}, \zeta^{N'} \} = \delta_{M'}^{N'}, \quad
[ \zeta_M, \partial^N \} = \delta_M^N $$

\subsection{ Superspin}

So far we have considered just coordinates, sufficient to define
scalar superfields.  For more general representations we may
want to consider superfields with indices, and thus ``superspin"
(in addition to the spin that comes from expanding the 
superfields in the anticommuting coordinates).
In terms of the supertwistor transform, these correspond to
\bea
\Phi_{M...}{}^{N'...} (w) = 
\int d\zeta\ exp \left(w_{M'}{}^N \zeta_N \zeta^{M'}\right) 
\zeta_M ... \zeta^{N'}... \chi (\zeta) +h.c. ~~.
\eea
The supertwistor form of the superconformal generators
is then the same as for the scalar superfield given above.

These representations follow from the constraint approach by the modification
\bea
 D_A{}^B = s_A{}^B, \quad D_{A'}{}^{B'} = s_{A'}{}^{B'}, \quad D_{A'}{}^B = 0 
 \eea
where the $s$'s are matrix representations of G$_0$.
The extra field equations involving the now nonvanishing 
$D_A{}^B$ and $D_{A'}{}^{B'}$ then fix the only representations
as $\Phi_{A...}{}^{B'...}$, where
\bea
 \Phi_{A...}{}^{B'...} = u_A{}^M ... u_{N'}{}^{B'} ... \Phi_{M...}{}^{N'...} (w) ~~.\eea

Off shell, we can consider more general representations.
For example, for the defining representations of the GL groups
we are considering, for the superfields $\Phi_{A'}$ and $\Phi^{A'}$
we still have 
\bea
 0 = D_{\cal A}{}^B = z_{\cal A}{}^{\cal M}\partial_{\cal M}{}^B
\quad\Rightarrow\quad \partial_{\cal M}{}^A = 0 
\eea
but are free to use $z_{A'}{}^{M'}=u_{A'}{}^{M'}$ to define
\bea
 \Phi_{A'} = u_{A'}{}^{M'}\Phi_{M'}, \quad
\Phi^{A'} = \Phi^{M'}u_{M'}{}^{A'} ~~.
\eea
The remaining constraint $D_{A'}{}^{B'} = s_{A'}{}^{B'}$ then
determines that $\Phi_{M'}$ and $\Phi^{M'}$ depend only on $w$.
We can apply a similar construction to $\Phi_A$ and $\Phi^A$ if
we first note that we can also write
\bea
 D_{\cal A}{}^{\cal B} = - z_{\cal M}{}^{\cal B} 
{\partial\over\partial z_{\cal M}{}^{\cal A}} \eea
to show
\bea
 D_{A'}{}^{\cal B} = 0 \quad\Rightarrow\quad 
{\partial\over\partial z_{\cal M}{}^{A'}} = 0 \eea
and thus use $z_M{}^A=u_M{}^A$ to define
\bea
 \Phi^A = \Phi^M u_M{}^A, \quad
\Phi_A = u_A{}^M\Phi_M \eea
where $\Phi^A$ and $\Phi_A$ also depend only on $w$.
(Note also that $\Phi^{A'}\Phi_{A'}=\Phi^{M'}\Phi_{M'}$
and $\Phi^A\Phi_A=\Phi^M\Phi_M$.)
These relations agree with the on-shell ones for the allowed cases.

Another way to derive the modification to the $D$'s is from
the modified $G$'s:  Starting with the supertwistor
expressions above, we find in terms of $\Phi_{M...}{}^{N'...} (w)$:
\bea
 G_M{}^{N'} = \partial_M{}^{N'}, \quad
G_M{}^N = w_{P'}{}^N\partial_M{}^{P'} +s_M{}^N, \quad
G_{M'}{}^{N'} = -w_{M'}{}^P\partial_P{}^{N'} +s_{M'}{}^{N'}, \eea
\bea
 G_{M'}{}^N = -w_{M'}{}^P w_{Q'}{}^N\partial_P{}^{Q'}
-w_{M'}{}^P s_P{}^N +s_{M'}{}^{P'}w_{P'}{}^N \eea
where these $s$'s act on the $\cal M$ indices of the fields that arise from
the supertwistor transform.
(Since these symmetries are global, they take the same form
on shell and off.)
The original field equations now also imply (from the new terms in $G$)
\bea
 s_{[M}{}^N \partial_{P)}{}^{Q'} = s_{M'}{}^{[N'}\partial_P{}^{Q')} = 0 \eea
which restrict the indices to those following from supertwistors.
Multiplying the $G$'s by $z$'s to ``flatten" the indices, we then find
for the $D$'s, when acting on constrained superfields (but either on or off
shell),
\bea
 D_{A'}{}^B = 0 \eea
\bea
 D_A{}^B = u_A{}^M s_M{}^N u_N{}^B \equiv s_A{}^B, \quad
D_{A'}{}^{B'} = u_{A'}{}^{M'} s_{M'}{}^{N'} u_{N'}{}^{B'} \equiv s_{A'}{}^{B'} \eea
\bea
 D_A{}^{B'} = u_A{}^M u_{N'}{}^{B'} \partial_M{}^{N'}
-s_A{}^C v_C{}^{B'} +v_A{}^{C'}s_{C'}{}^{B'} ~~.\eea

\section{ Projective lightcone  limit}

The projective lightcone limit of AdS$_5\otimes$S$^5$ \cite{[3]} can be defined as
\bea
 z_A{}^{\cal M} \rightarrow \sqrt R z_A{}^{\cal M}, \quad
z_{A'}{}^{\cal M} \rightarrow {1\over \sqrt R}z_{A'}{}^{\cal M}; \quad\quad
R \rightarrow 0 \eea
which preserves the superconformal group, but not the isotropy group.
(Effectively, the $\cal M$ indices are left invariant, while
the $A$ and $A'$ indices are scaled oppositely.)
Actions and metrics can be written in terms of the superconformally
invariant (but not isotropic) differential forms
\bea
 J_{\cal A}{}^{\cal B} = (dz_{\cal A}{}^{\cal M})z_{\cal M}{}^{\cal B} 
 \eea
whose limit is
\bea
 J_A{}^{B'} \rightarrow R  J_A{}^{B'}, \quad
J_A{}^B \rightarrow J_A{}^B, \quad
J_{A'}{}^{B'} \rightarrow J_{A'}{}^{B'}, \quad
J_{A'}{}^B \rightarrow {1\over R} J_{A'}{}^B ~~.\eea
In this limit, the leading contribution to actions and metrics is given
by just $J_{A'}{}^B$.  In terms of the above change of variables
(for which the factored matrix expression is convenient),
\bea
 J_{A'}{}^B = u_{A'}{}^{M'}(dw_{M'}{}^{N})u_N{}^B ~~.\eea

To relate to previous results, we first examine the projective
lightcone limit for just AdS$_5$.  Corresponding to the superstring, we use
the covering groups GL(4)/Sp(4)GL(1), rather than SO(3,3)/SO(3,2).
The limit is defined \cite{[3]} by scaling the coordinate ``$x_0$" (which
vanishes on the boundary) with respect to the remaining coordinates.
In spinor (covering group) notation, this is the special case $A=\alpha$, $A'=\dot\alpha$ of the above.  
The limit leaves GL(4) intact, while contracting Sp(4) to the Poincar\'e group.  
The metric of AdS$_5$ in terms of the GL(4)-invariant forms is
\bea
 ds^2 = R^2 J_{\cal A}{}^{\cal B}J_{\cal C}{}^{\cal D}
\Omega^{\cal AC}\Omega_{\cal BD} \eea
using the Sp(4) metric $\Omega_{\cal AB}$ and its inverse.
The limiting form is then
\bea
 ds^2 \rightarrow J_{\dot\alpha}{}^\beta J_{\dot\gamma}{}^\delta
C^{\dot\alpha\dot\gamma}C_{\beta\delta} \eea
where
\bea
 J_{\dot\alpha}{}^\beta = u_{\dot\alpha}{}^{\dot\mu}
(dx_{\dot\mu}{}^\nu) u_\nu{}^\beta \eea
\bea
 \Rightarrow\quad ds^2 = {dx^2\over (x_0)^2}; \quad\quad
dx^2 = dx_{\dot\alpha}{}^\beta dx_{\dot\gamma}{}^\delta
C^{\dot\alpha\dot\gamma}C_{\beta\delta}, 
\quad x_0 = |\textstyle{1\over 2}
u_\alpha{}^\mu u_\beta{}^\nu C^{\alpha\beta} C_{\mu\nu} | \eea
and we have used the fact that after Wick rotation complex conjugation
relates $u'$ to $u^{-1}$ (not $u$).

The Green-Schwarz action for  AdS$_5\otimes$S$^5$ \cite{[20]}
can be expressed as quadratic \cite{[21]}
in the above currents $J_{\cal A}{}^{\cal B}$,
with an overall factor of $R^2$.
Upon taking the limit, where only $J_{A'}{}^B$ survives,
the only term remaining is again
\bea
 L \rightarrow J_{\dot\alpha}{}^\beta J_{\dot\gamma}{}^\delta
C^{\dot\alpha\dot\gamma}C_{\beta\delta} 
\eea
but now
\bea
 J_{\dot\alpha}{}^\beta = u_{\dot\alpha}{}^{M'}
(dw_{M'}{}^N) u_N{}^\beta 
\eea
Plugging into $L$, the SL(2) parts of $u$ and $u'$ are again canceled,
and the GL(1) parts again combine into $x_0$, leaving only the fermions.
Explicitly, we can write
\bea
 u_M{}^\alpha = ( \delta_\mu^\nu, \vartheta_m{}^\nu ) u_\nu{}^\alpha, 
\quad u_{\dot\alpha}{}^{M'} = u_{\dot\alpha}{}^{\dot\nu} 
( \delta_{\dot\nu}{}^{\dot\mu}, \bar\vartheta_{\dot\nu}{}^{m'} ) 
\eea
We then have
\bea
 L ={( J_{\dot\mu}{}^\nu)^2 \over (x_0)^2}, \quad\quad
J_{\dot\mu}{}^\nu = dx_{\dot\mu}{}^\nu 
+\bar{\vartheta}_{\dot\mu}{}^{m'}d\theta_{m'}{}^\nu
+(d\bar{\theta}_{\dot\mu}{}^m)\vartheta_m{}^\nu
+\bar{\vartheta}_{\dot\mu}{}^{m'}(dy_{m'}{}^n)\vartheta_n{}^\nu ~~.\label{312}\eea

Note that our definition of the projective lightcone limit for the
supersymmetric case treats S$^5$ in a similar way to AdS$_5$,
in that a five-dimensional space of constant curvature 
has been contracted to a flat four-dimensional space.
Previously we contracted the sphere to a point, as the only
obvious contraction that preserved the internal symmetry.
This new contraction actually produces a complex internal space,
but such spaces are standard in harmonic and projective approaches
to extended supersymmetry, and reality properties of superfields
are preserved by the usual generalization of complex conjugation
for fields whose coordinates have been Wick rotated.

The usual quantization of this string action is expected to be
difficult, as it was before the limit, particularly since the
Wess-Zumino term has dropped out (but will contribute to
corrections higher-order in $R$).  Instead, we want to consider
random-lattice quantization of this action \cite{[3]}, which effectively
means that it is treated as the action for a superparton of the
superstring (so dropping the Wess-Zumino term is actually an
advantage).  In the following section we consider this and
related superparticle actions.

\section{ Proposed actions for N=4 YM}

\subsection{ Holographic action}

We now apply the above action to the superparticle.
In addition to the coordinates expected for projective superspace
(in this case,
GL(4|4)/GL(2|2)$^2$+), we have found ``Lagrange
multipliers" $x_0$ and $\vartheta$.  The former appear with
derivatives,
the latter without:  By the usual canonical quantization,
this means the fields depend only on the projective superspace
coordinates.  ($\partial/\partial$(Lagrange multiplier)=0.) 
However, there are still the usual second-class constraints:
In fact, without the $y$ terms, this is just the usual 
Casalbuoni-Brink-Schwarz action for the superparticle,
up to a redefinition of $x$.  (We identify $(x_0)^2$ as the
worldline metric; the conformal construction has guaranteed
its positivity.)  So, this action is a generalization of that to
projective superspace.  

In particular, the case of chiral
superspace (n=0) has no $y$'s, and this treatment of the usual
superparticle action merely suggests the separation of first-
and second-class constraints:  The usual $\bar\theta$'s are
now exactly the $\bar\vartheta$'s, so the restriction to chiral
superspace follows from interpretation of the $\bar\vartheta$'s
as Lagrange multipliers.  The redefinition of $x$ that led to
their appearing without derivatives was just the usual (complex)
coordinate transformation to the chiral representation.
(Of course, antichiral superspace, n=N, is also equivalent to
the usual superparticle, after the opposite transformation.)

In the general case, if we make the redefinitions 
(in matrix notation)
\bea
 \theta \rightarrow \theta -y \vartheta, \quad
\bar\theta \rightarrow \bar\theta - \bar\vartheta y, \quad
x \rightarrow x +\bar\vartheta y \vartheta \eea
we find
\bea
 J \rightarrow dx +\bar\vartheta d\theta 
+(d\bar\theta) \vartheta \eea
i.e., the only effect is to remove $y$.  Furthermore,
the redefinition
\bea
 x \rightarrow x -\textstyle{1\over 2}
(\bar\vartheta\theta +\bar\theta\vartheta) \eea
gives the usual superparticle action
(with the usual reality properties), since
\bea
 J \rightarrow dx +\textstyle{1\over 2}
(\bar\vartheta \on\leftrightarrow\partial \theta 
-\bar\theta \on\leftrightarrow\partial \vartheta) ~~.\eea

The addition of pure gauge degrees of freedom
(in this case, $y$) to the classical mechanics formulation of
a theory, and the resulting nonminimal terms to the
first-quantized Becchi-Rouet-Stora-Tyutin operator,
is often required for covariant second-quantization,
and thus for the gauge- and Lorentz-covariant
formulation of the classical field theory \cite{[22]}.
In the harmonic and projective approaches to superspace,
the addition of the internal coordinates for
CPT self-conjugate multiplets is necessary to introduce
the infinite number of auxiliary fields required for a
manifestly supersymmetric formulation.
On the other hand, in terms of on-shell lightcone
superfields, all formulations are the same, up to
trivial Fourier transformation in some of 
the lightcone $\theta$'s (which trades a $\theta$ for a
$\bar\theta$), since the on-shell field content is unique.
This also can be seen covariantly from the supertwistor 
transform given above:  In terms of the same supertwistor
field $\chi(\zeta)$, reassigning which of the anticommuting
components belong to $\zeta_M$ and which to $\zeta^{M'}$
results in a different choice of coordinates $w$ in $\Phi(w)$.

The constraints following from a canonical
analysis of the system \bref{312} are
\bea
(D_\mu{}^{\dot{\nu}})^2=\left(\frac{\partial}{\partial x_{\dot{\nu}}{}^\mu}\right)^2
=0\label{45}
\eea
which comes from  variation of $x^0$, as well as
the usual mixed first- and second-class constraints
that follow from the definition of the canonical fermionic momenta,
\bea
D_\mu{}^{n'}=\frac{\partial}{\partial \theta_{n'}{}^\mu}+\bar{\vartheta}_{\dot{\nu}}{}^{n'}
\frac{\partial}{\partial x_{\dot{\nu}}{}^\mu}=0,
\quad 
D_m{}^{\dot{\nu}}=\frac{\partial}{\partial \bar{\theta}_{\dot{\nu}}{}^m}
-\frac{\partial}{\partial x_{\dot{\nu}}{}^\mu}
\vartheta_m{}^{\mu}
=0,\label{46}\eea
\bea
D_\mu{}^{n}=\frac{\partial}{\partial {\vartheta}_{n}{}^\mu}=0,\quad
D_{m'}{}^{\dot{\nu}}=\frac{\partial}{\partial \bar{\vartheta}_{\dot{\nu}}{}^{m'}}=0\label{47}
\eea
and the one following from the definition of the canonical conjugate of $y$,
\bea
\frac{\partial}{\partial y_{n'}{}^m}+\bar{\vartheta}_{\dot{\mu}}{}^{n'}
\frac{\partial}{\partial x_{\dot{\mu}}{}^\nu}\vartheta_m{}^\nu=0\label{yy}
~~.
\eea
The primary constraint \bref{yy} can be recast 
into the first-class constraint
\bea
D_m{}^{n'}
=\frac{\partial}{\partial y_{n'}{}^m}+\bar{\vartheta}_{\dot{\mu}}{}^{n'}
\frac{\partial}{\partial x_{\dot{\mu}}{}^\nu}\vartheta_m{}^\nu
-\vartheta_m{}^\mu D_\mu{}^{n'}-D_m{}^{\dot{\nu}}\bar{\vartheta}_{\dot{\nu}}{}^{n'}=0~~
\label{48},
\eea
since $y$ is pure gauge.
These constraints satisfy the usual relations
$\{D_\mu{}^{n'},D_{m'}{}^{\dot{\nu}}\}=\delta_{m'}^{n'}D_\mu{}^{\dot{\nu}}$
and  $\{D_m{}^{\dot{\nu}},D_{\mu}{}^{n}\}=-\delta_{m}^{n}D_\mu{}^{\dot{\nu}}$,
expressing their second-class nature.  They can be reduced to the
first-class constraints $D_{[M}{}^{[M'}D_{N)}{}^{N')}$, 
$D_\mu{}^{n}$, and $D_{m'}{}^{\dot{\nu}} $.
The linear constraints, $D_\mu{}^{n}$ and $D_{m'}{}^{\dot{\nu}} $,
are imposed as isotropy constraints, 
implying that the superfields are
independent of $\vartheta$ and $\bar{\vartheta}$.
The quadratic first-class constraints are the
projective superspace field equations \bref{220}.
They include the generator of $\kappa$
symmetry transformations just as in the Casalbuoni-Brink-Schwarz action. 
The constraints \bref{45}-\bref{48}
are consistent with the
superconformal field equations \bref{226}.  
 
\subsection{ Dual Hamiltonian action}

This action, and its relation to projective superspace,
suggests alternative actions that do not suffer from
second-class constraints.  First, noting the similarity
(``duality" in the differential geometry sense) between
the expressions for the forms $J_{A'}{}^B$ and the
covariant derivatives $D_A{}^{B'}$, we are led to propose
the action
\bea
 L' = \dot w_{M'}{}^N p_N{}^{M'} -(D_\alpha{}^{\dot\beta})^2,
\quad\quad D_\alpha{}^{\dot\beta} = 
u_\alpha{}^M p_M{}^{N'}u_{N'}{}^{\dot\beta} ~~.
\eea
Separating the $u$'s in a way dual to the previous,
\bea
 u_\alpha{}^M = u_\alpha{}^\nu ( \delta_\nu^\mu , \phi_\nu{}^m ),
\quad u_{M'}{}^{\dot\alpha} = ( \delta_{\dot\mu}^{\dot\nu}, 
\bar\phi_{m'}{}^{\dot\nu} ) u_{\dot\nu}{}^{\dot\alpha} 
\eea
we have
\bea
 L' = \dot w_{M'}{}^N p_N{}^{M'} -(x_0)^2 (D_\mu{}^{\dot\nu})^2,
\quad\quad D_\mu{}^{\dot\nu} = 
p_\mu{}^{\dot\nu} +\phi_\mu{}^m \bar\pi_m{}^{\dot\nu}
+\pi_\mu{}^{m'}\bar\phi_{m'}{}^{\dot\nu}
+\phi_\mu{}^m T_m{}^{n'} \bar\phi_{n'}{}^{\dot\nu} \nn\\
\eea
where $p,\pi,T$ are conjugate to $x,\theta,y$.

An expansion in $\phi$, as follows from the canonical
$\partial/\partial\phi=0$,
leads exactly to the superconformal equations of motion \bref{226}.

\subsection{ Quadratic multiplier action}

Another suggested alternative follows from noticing that the
Lagrange multipliers $u$ appear quartically in the previous
actions, although there can be some simplification because
they are multipled directly as $u^2$ and $u'^2$.  Then a
manifestly GL(2|n)GL(2|N-n)-covariant Hamiltonian action
can be written as
\bea
 L'' = \dot w_{M'}{}^N p_N{}^{M'} 
-e^{[MN)}e_{[M'N')}p_M{}^{M'}p_N{}^{N'} ~~.
\eea
Again the positivity of 
$x_0=|\textstyle{1\over 2}e^{\mu\nu}C_{\mu\nu}|$ 
is guaranteed.  Many of the terms are similar to those in
$L'$, but now there are no terms higher than quartic.

\section{ Conclusions}

We have shown that N=4 super Yang-Mills is the
underlying theory of the random lattice approach
to quantization of the superstring about the
AdS$_5\otimes$S$^5$ background.
The addition of internal coordinates to the
Casalbuoni-Brink-Schwarz action implies the use of
projective superspace for off-shell superfields.
While the physical content of the action is unchanged,
the introduction of the auxiliary internal variables
may allow a simpler treatment of covariant 
first-quantization, just as it does for second-quantization.
Effectively, the new coordinates are gauge degrees of freedom
at the first-quantized level, but introduce auxiliary fields
at the second-quantized level, useful for covariant treatments.
Alternatively, one of the proposed related actions might 
prove useful in avoiding second-class constraints.\par
\vskip 6mm\noindent
{\bf Acknowledgments}

 We thank Tirthabir Biswas and Martin Ro\v cek for discussions on
harmonic and projective superspace.
W.S. was supported in part by
National Science Foundation Grant No.\ PHY-0098527.
M.H. acknowledges CNYITP at Stony Brook for warm hospitality
and KEK for support.

\end{document}